**Band gap formation theory: An alternative to the Bragg diffraction model**

Koichi Kajiyama[1,**]

[1]New Industry Creation Hatchery Center, Tohoku University, Sendai, Japan

**ABSTRACT**. The band gap, a key concept in solid-state physics, is traditionally explained by the Bragg diffraction of electron waves in the periodic potential of a crystal. Although widely accepted, this framework raises fundamental issues in one-dimensional systems, where Bragg diffraction—which requires multidirectional wave interactions—reduces to simple interference, thus failing to explain band gap formation. In this paper, we introduce an alternative theory that does not rely on Bragg reflection. Using the Schrödinger equation for Bloch waves, we consider the crystal lattice as a discrete set of observation points. This discreteness introduces a sampling-like constraint analogous to the Nyquist frequency in signal processing. We show that when the electron wavenumber changes under a periodic potential while the lattice spacing remains fixed, a band gap naturally emerges as a sampling effect. By constructing an energy diagram that incorporates this effect, we reveal that the band gap originates from both the wavenumber change and the role of the lattice as discrete samplers, leading the energy curve to exhibit translational and mirror symmetries with respect to the Nyquist wavenumber. This approach provides a novel, physically grounded explanation of band gap formation and can be naturally extended to higher dimensions.

## I. INTRODUCTION.

The band gap is a fundamental concept in solid-state physics that plays a central role in the electrical, optical, and thermal properties of materials. Traditionally, the formation of band gaps has been explained by the concept of Bragg diffraction, where electron waves are reflected by the periodic potential of the crystal lattice, resulting in standing-wave patterns and energy gaps. This explanation is widely accepted and appears in most standard textbooks [1], particularly those involving the Kronig–Penney model in one-dimensional (1D) systems. However, a critical examination of this view reveals inconsistencies when it is strictly applied to 1D systems. Bragg diffraction inherently involves wave interactions across directions and angles, phenomena that are inherently two- or three-dimensional. In a truly 1D lattice, wave reflection, as described by Bragg diffraction, becomes indistinguishable from simple interference, questioning its role as the primary cause of band gap formation in such lattices. This insight served as the starting point for the present study. As a result of this research, we propose an alternative theoretical interpretation of band gap formation that does not rely on Bragg diffraction. Instead, we investigate a 1D crystal model based on the Schrödinger equation and Bloch's theorem, focusing on the wave properties of electrons in the presence of a periodic potential. A key point of our approach is the discrete nature of the lattice points considered as observation points for the wavefunction, introducing a fundamental limit akin to the Nyquist frequency in signal processing.

When a periodic potential is present in the lattice, the wavelength (or wavenumber) of the propagating wave changes while the lattice spacing remains constant. This was found to cause a sampling-like effect, in which the discrete structure of the lattice imposes a limit on how the wave can be observed or reconstructed. This concept mirrors the principles of the Nyquist–Shannon sampling theorem [2, 3], where frequency and time are analogous to the wavenumber and spatial position, respectively.

## II. BLOCH WAVE IN STEADY STATE

### A. Carrier and modulation components in a steady state

$$\psi_k(x) = u_k(x)e^{ikx}, \quad u_k(x + a) = u_k(x) \quad (1)$$

This expression, known as the Bloch wave, underpins the electronic band structure in periodic systems, where $a$ is the lattice constant. In an isolated crystal, Bloch waves are stationary. The function $u(x)$, localized at lattice points, is approximated as

[*]Corresponding author: koichi.kajiyama.d3@tohoku.ac.jp

$$u(x) \approx \frac{1}{a}\sum_{n=-nmax}^{n=nmax} \cos\left(\frac{2\pi}{a}\right)x$$
$$= \frac{1}{a}\sum_{n=-nmax}^{n=nmax} \cos(k_n x) \quad (2)$$

The amplitude peaks sharply at regular intervals, reflecting strong localization at the lattice points as shown in Fig. 1. This reflects the periodic and symmetric nature of $u_k(x)$.

The modulation wave is also stationary, with boundary conditions determined by the crystal length, which is substantially larger than the lattice spacing. Thus, $k$ can be treated as quasi-continuous, and because the system is in a steady state, $e^{ikx}$ can be replaced by $cos(kx)$ as follows:

$$\phi_{mod}(k,x) = cos(kx) \quad (3)$$

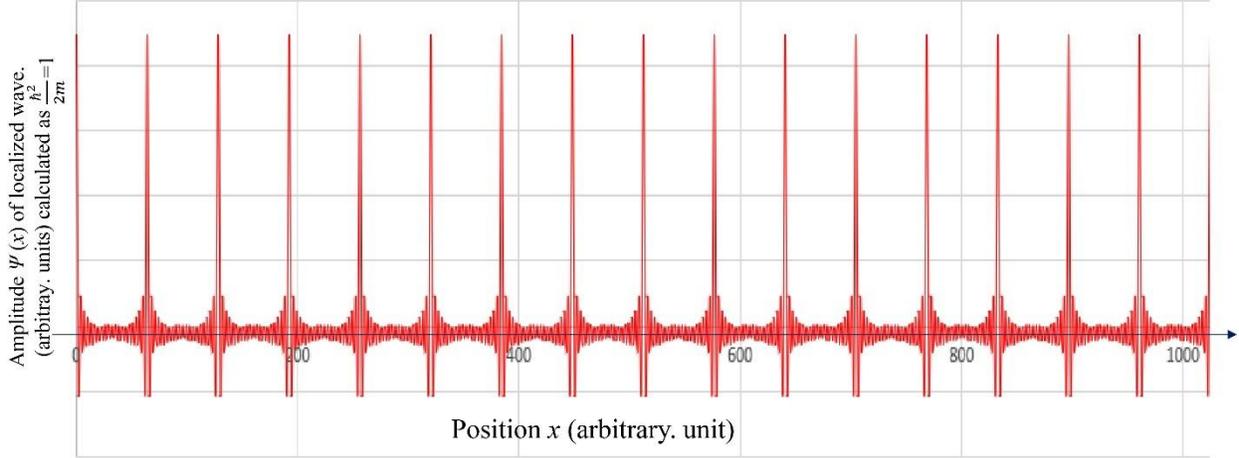

Fig. 1. Spatial profile of localized Bloch wave calculated under the unit system, where $\frac{\hbar^2}{2m}=1$.
The horizontal axis represents the position $x$, and the vertical axis shows the amplitude $\psi(x)$ in arbitrary units. Sharp periodic peaks indicate strong localization at discrete lattice points.

Consequently, the Bloch wave in a steady state becomes

$$\phi_{Bloch}(k,x) = u_k(x)\,\phi_{mod}(k,x), \text{ where } u_k(x)= \frac{1}{a}\sum_{n=-nmax}^{n=nmax} \cos(k_n x) \quad (4)$$

### B. Translational symmetry of Bloch waves in reciprocal space

Fig. 2 illustrates typical Bloch waves modulated by two functions $\phi_{mod}(k,x)$ with wavenumbers $k$ that differ by $2\pi/a$, which results in identical Bloch waves owing to the periodicity in the reciprocal space. Equation (4) can be rewritten by applying the trigonometric identity shown in Eq. (5).

$$\phi_{Bloch}(k,x) = \frac{1}{a}\sum_{n=-nmax}^{n=nmax}(\cos(k_n x)\cos(kx))$$
$$= \frac{1}{2a}\sum_{n=-nmax}^{n=nmax}(\cos(k_n+k)x + \cos(k_n-k)x) \quad (5)$$

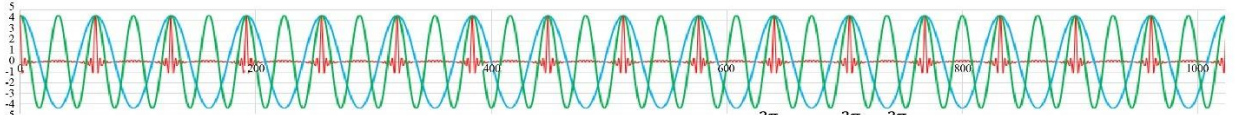
Fig.2-1. No modulated Bloch wave for two modulation waves. $k_1 = \frac{2\pi}{a}, k_2 = \frac{2\pi}{a} + \frac{2\pi}{a}$.

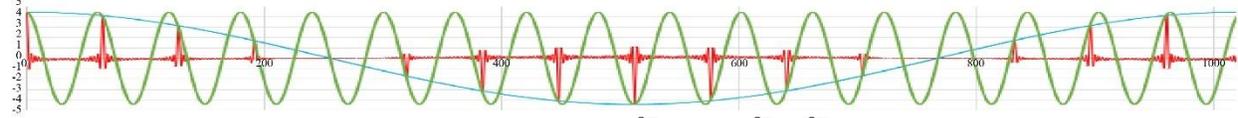
Fig.2-2. Wavenumber of $k_1$ is rather small to $k_2$. $k_1 = \frac{2\pi}{1024}$, $k_2 = \frac{2\pi}{1024} + \frac{2\pi}{a}$.

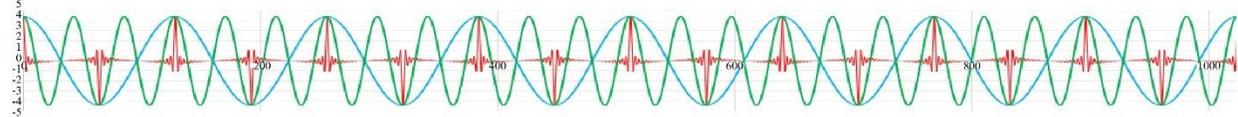
Fig.2-3. Wavenumber of $k_1$ is half of $\frac{2\pi}{a}$. $k_1 = \frac{\pi}{a}$, $k_2 = \frac{\pi}{a} + \frac{2\pi}{a}$.

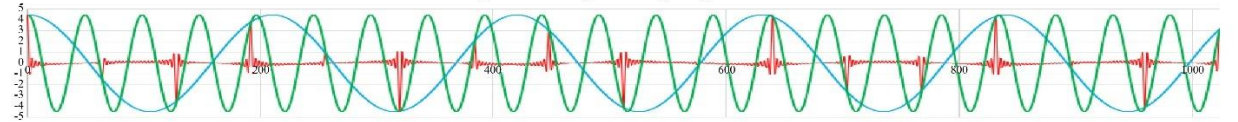
Fig.2-4. Interference is complicated, but still periodical. $k_1 = \frac{2\pi}{210}$ $k_2 = \frac{2\pi}{210} + \frac{2\pi}{a}$.

Fig. 2. Modulated waves differing by $\frac{2\pi}{a}$ in $k$ represent the same Bloch wave.
(Bloch waves for $k_1$ and $k_2$ are shown in blue and green, respectively.)
The shape of a Bloch wave is determined by the interference between the discretely and periodically spaced lattice points and modulation wave $\phi_{mod}(k, x)$. Because the lattice points are discrete, two modulation waves whose wavenumbers differ by 2π/a produce the same Bloch wave pattern. This gives rise to the translational symmetry in reciprocal space, as described in Section II.B.

This transformation decomposes the product of two cosine terms into a sum of two cosine functions with wavenumbers $k_n \pm k$. The result reveals that each term in the Bloch wave expression corresponds to the superposition of waves with symmetric wavenumbers around $k_n$, as follows:

$$\phi_{Bloch}(k,x) = \frac{1}{2a} \sum_{n=-nmax}^{n=nmax} \{\cos(k_n + k)x + \cos(k_n - k)x\} \quad k_n = \frac{2\pi n}{a} \quad (6)$$

This decomposition is essential for understanding the underlying symmetry in reciprocal space. By setting the spatial coordinate to $x = a$, the expression $\cos\{(k_n + k)a\}$ reduces to $\cos(ka)$, indicating that the Bloch wave remains invariant under changes in $n$, since $k_n = \frac{2\pi n}{a}$. This invariance demonstrates the translational symmetry of the Bloch wave with respect to the lattice spacing $a$, which is evident both from the mathematical formulation and from the behavior depicted in Fig. 2.

### C. Linear symmetry with respect to Nyquist wavenumber

In addition, linear symmetry exists with respect to the Nyquist wavenumber and is given by

$$k_{Ny} = \frac{2\pi}{2a} = \frac{\pi}{a} \quad \text{or more generally} \quad k_{Ny} = \frac{\pi}{a} + \frac{2\pi n}{a} \quad (7)$$

This relationship is clearly visualized in Fig. 3. For more detailed information on the Nyquist frequency $\omega_{Ny}$, please refer to [2] and [3].

The Nyquist frequency originates from signal processing in communication systems, where the critical sampling rate necessary to capture a signal without aliasing is defined. This concept can also be applied to the energy diagram of a Bloch wave by substituting frequency $\omega$ and time $t$ with wavenumber $k$ and position $x$. Here, we compare the Bloch waves with the wavenumbers $k_{Ny} \pm \Delta k$, which are symmetrically located with respect to $k_{Ny}$. To further illustrate the symmetry around the Nyquist wavenumber $k_{Ny}$, we examine the modulated function $\cos(kx)$ at $k = k_{Ny} \pm \Delta k$, where $k_{Ny} = \frac{\pi}{a}$. Applying the trigonometric identity for angle addition yields the following equation:

$$\cos\left\{\left(\frac{\pi}{a} \pm \Delta k\right)x\right\} = \cos\left(\frac{\pi}{a}x\right)\cos(\Delta kx) \mp \sin\left(\frac{\pi}{a}x\right)\sin(\Delta kx) \quad (8)$$

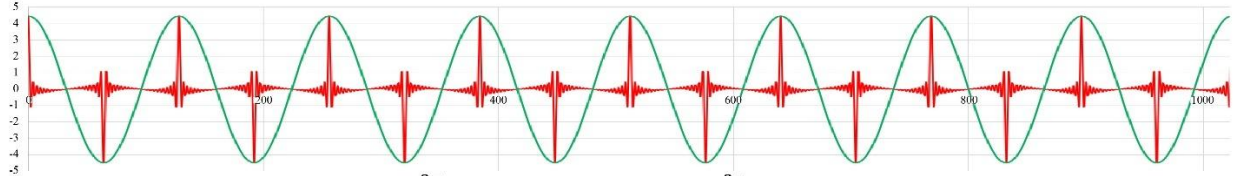
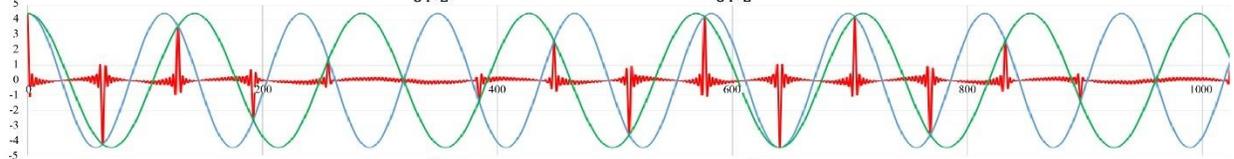
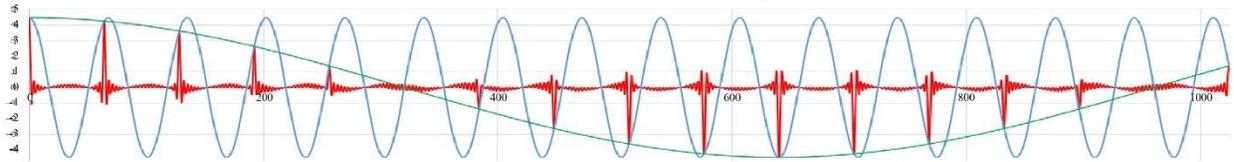

Fig.3-1. Blue line: $k_1 = \frac{2\pi}{64*2} * (1 + 0)$, Green line: $k_2 = \frac{2\pi}{64*2} * (1 - 0)$,

Fig.3-2. Blue line: $k_1 = \frac{2\pi}{64*2} * (1 + 0.1)$, Green line: $k_2 = \frac{2\pi}{64*2} * (1 - 0.1)$,

Fig.3-3. Blue line: $k_1 = \frac{2\pi}{64*2} * (1 + 0.9)$, Green line: $k_2 = \frac{2\pi}{64*2} * (1 - 0.9)$

Fig. 3. Line Symmetry of Bloch Waves with Respect to the Nyquist Wavenumber.
($k_{Ny} = \frac{\pi}{a}$). $a$: lattice distance.
Due to the line symmetry about the Nyquist wavenumber, modulation waves with wavenumbers $k = k_{Ny} \pm \Delta k$ result in the same Bloch wave shape.

When these values ($x = a$) are substituted, the expression simplifies to

$$cos\{(k_{Ny} \pm \Delta k)a\} = -cos(\Delta k \times a) \quad (9)$$

In this context, the Nyquist frequency may be more appropriately referred to as the Nyquist wavenumber. A more detailed analysis is provided below. The Nyquist wavenumber is represented by the dashed line at the boundary of the Brillouin zone shown in Fig. 4. This implies that even if the wavenumber of a Bloch wave varies continuously, the waveform changes discontinuously at the Nyquist wavenumber, as illustrated in Fig. 3.

This result demonstrates that the modulated function is symmetric around $k_{Ny}$ and that at $x = a$, both $k_{Ny} + \Delta k$ and $k_{Ny} - \Delta k$ yield identical values. This symmetry is fundamental to the periodicity in reciprocal space and plays a key role in the formation of the band gap. The modulation wave $cos(kx)$ in a Bloch wave corresponds to an alternating electrical waveform $cos(\omega t)$. Here, the lattice spacing $a = \Delta x = 2\pi/k$ corresponds to the sampling time interval $\Delta t = 2\pi/\omega$. When the wave is sampled at regular spatial intervals, as in time sampling, the resulting pattern resembles the periodic lattice shown in Fig. 3. The only difference is that time is unidirectional, whereas space is symmetric.

### III. ROLE OF LATTICE POINTS IN THE ABSENCE OF POTENTIAL OR SCATTERING

What is the true physical role of the lattice points that neither scatter Bloch waves nor possess any potential?

**A. Crystal lattice as a set of observation points**

Although a full treatment of the measurement problem lies beyond the scope of this study, the present formulation highlights that the discrete nature of the observation points—regardless of whether they interact directly with the system—may influence the observed symmetry and energy characteristics of the Bloch wave, similar to the sampling effects in classical signal analysis.

**B. Relationship between the observation points and uncertainty**

This influence arises from the fundamental uncertainty associated with sampling; in fact, finite spacing $a = \Delta x$ implies that the information between the lattice points is inaccessible, leading to discretized reciprocal spacing $\Delta k = 2\pi/a$ and an associated uncertainty relation $\Delta x \Delta k = 2\pi$. Although $x$ and $k$ are connected through this relation, these two variables remain independent because the presence of a potential can modify them independently, as discussed in Section IV.

# IV. ENERGY DIAGRAM IN THE ABSENCE OF A PERIODIC POTENTIAL

## A. Spectral composition and Brillouin zones

In the absence of a periodic potential, the Schrödinger equation of a Bloch wave is expressed as

$$-\frac{\hbar^2}{2m}\frac{d^2}{dx^2}\phi_{Bloch}(k,x) = \frac{\hbar^2 k^2}{2m}\phi_{Bloch}(k,x) = E_0\phi_{Bloch}(k,x) \quad (10)$$

Fig. 4 shows the energy diagram in the absence of a periodic potential. Each Brillouin zone is shaded, with the Nyquist wavenumbers $k_{Ny} = \frac{\pi}{a} + \frac{2\pi n}{a}$ indicated by the dashed dotted line. This diagram, which is representative of Bloch waves, comprises multiple quadratic curves of the form $E = (\hbar^2 k^2)/2m$ arranged according to translational symmetry. The energy diagram for the case without potential has already been established (see Fig. 4). This diagram reveals both translational symmetry at $k = nka$ and mirror symmetry at the Nyquist wavenumber $k = k_{Ny}$. Therefore, it is sufficient to consider only the range $-k_{Ny} \leq k \leq k_{Ny}$.

The mathematical formulation of this process is straightforward. According to the Schrödinger equation, the energy of a wavefunction is proportional to the square of the wavenumber. Consequently, the functions depicted in Fig. 4 are all quadratic functions with identical shapes owing to symmetry. These functions can be expressed as follows:

$$E_0(k) = \frac{\hbar^2}{2m}k^2 \text{ or more generally } E_0(k) = \frac{\hbar^2}{2m}(k_n \pm k)^2 \quad (11)$$

where $E_0(k)$ represents the energy in the absence of a potential. However, the focus of the present discussion is on the points where each curve intersects the $k = 0$ axis. These values, as expected, correspond to the second derivative of each term in Eq. (2), specifically the square of $k_n$. Starting from each point along $k = 0$ (marked with asterisks), it becomes evident, according to Eq. (4), that each component modulated by $cos(kx)$ splits into two sinusoidal waves, as expressed in Eq. (5). Therefore, instead of interpreting the diagram merely as a collection of quadratic curves, it is more insightful to view it as showing how each frequency component of the carrier wave splits under modulation. These components fold back in a line-symmetric manner at the Nyquist wavenumber and collectively form an energy curve that satisfies both mirror and translational symmetries.

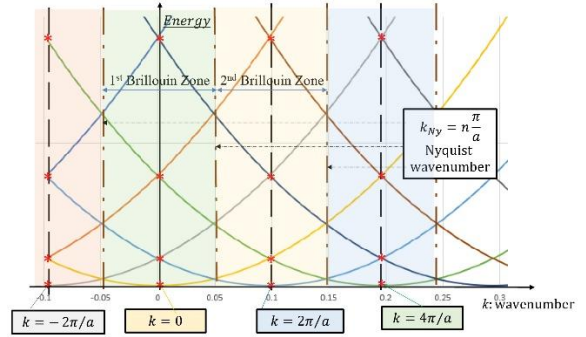

Fig. 4. Energy diagram of Bloch waves without periodic potential. $a = 64 : E_0(k) = (k_n + k)^2 : (k_n = n \times \frac{2\pi}{a})$

Reflecting both the translational and mirror symmetries of the Bloch wave, the energy diagram also exhibits these symmetries. A notable feature is the point at $k = 0$, where no modulation is present. In detail, the corresponding spectral component represents a Bloch wave strongly localized at the lattice points, as illustrated in Fig. 1. This waveform is determined solely by the lattice structure, unaffected by the potential and thus represents the fundamental form of the Bloch wave.

As will become clear in the next section, this interpretation of the energy diagram provides a more physically intuitive understanding of the band structure. Before introducing the effect of a periodic potential, we first clarify the energy dispersion of the Bloch wave in the free-electron case.

# V. QUANTIFICATION OF BAND GAP FORMATION UNDER A PERIODIC POTENTIAL

## A. Concept and procedure

Bloch waves under a periodic potential are typically explained by Bragg diffraction, which attributes the band gap to wave reflection within the crystal lattice. In contrast, we propose that the band gap arises from a change in the wavenumber of the modulation wave $\phi_{mod}(k,x)$ while the lattice spacing remains fixed. This phenomenon is analogous to filling a tank with water: although the physical dimensions of the tank do not change, the internal refractive index shifts from that of air to that of water. Consequently, the wavenumber of a light wave propagating through the tank changes accordingly, leading to a phase shift in the light.

As shown in Section IV, the effect of the periodic potential on the energy of a Bloch wave with respect to its wavenumber involves two key factors related to the formation of the band gap. First, the energy is proportional to the square of the wavenumber, which implies that the curvature of this quadratic curve represents the second derivative of the energy with respect to $k$. However, the multiple points of the Bloch wave along the vertical axis at $k = 0$ (marked with asterisks) correspond to the periodic lattice function $u(x)$, which remains unaffected by the potential. This is because $u(x)$ represents the inherent lattice structure, and at $k = 0$, the modulation wave becomes constant at a value of 1, without amplitude modulation.

These two conditions are essential for accurately describing the formation of the band gap. That is, the resulting energy function must be a quadratic curve constrained by the fixed lattice constant.

### B. Determining the curvature coefficient $B_{EC}$ for energy dispersion

Having established the baseline energy diagram for the case without potential, we now introduce the effect of a periodic potential. To reflect the influence of the periodic potential in the Schrödinger equation, we modify Eq. (10) as follows:

$$-\frac{\hbar^2}{2m}\frac{d^2}{dx^2}\phi_{Bloch}(k,x) = \left(\frac{\hbar^2 k^2}{2m} - V\right)\phi_{Bloch}(k,x) = E_p\phi_{Bloch}(k,x) \quad (12)$$

This differential equation represents the kinetic energy of a Bloch wave under a periodic potential $V$. Consequently, the kinetic energy in the presence of a periodic potential becomes $E_p = E_0 - V$. However, the crystal lattice remains completely unaffected.

Then, at $k = 0$, the Bloch wave represents the crystal lattice, and its waveform corresponds to that shown in Fig. 1. The energy reaches its minimum value at this point, which corresponds to the case of $n = 0$, resulting in $E = 0$, regardless of the presence or absence of a potential. Consequently, one fixed point of the quadratic function is established.

Fig. 5 illustrates a model based on a classical interpretation of Eq. (12). This model reveals that the influence of the potential is the greatest when the momentum reaches its maximum value. This may be regarded as a definition of the potential. Owing to the folding at the Nyquist wavenumber, the maximum kinetic energy for the case of $n = 0$ occurs at $k = k_{Ny}$.

In other words, the energy at $k = k_{Ny}$ is given by $E_p(k_{Ny}) = E_0(k_{Ny}) - V$, which represents the maximum reduction in kinetic energy owing to the periodic potential.

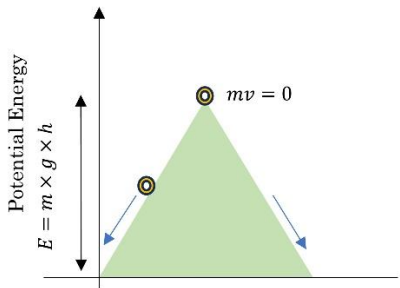

Fig.5.1. Momentum energy and Potential.

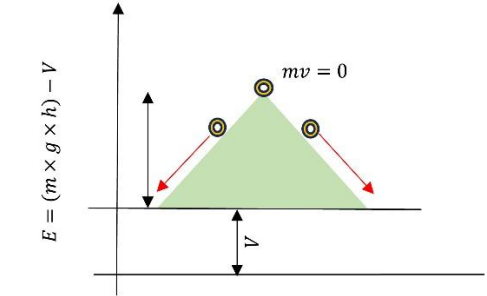

Fig.5.2. The energy decreased by a potential energy $V$.

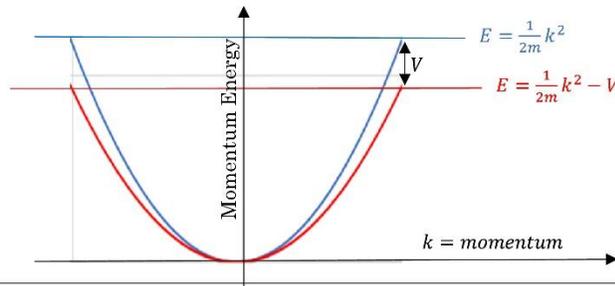

Fig.5.3. Relation between a Potential energy and Momentum energy in a classical model.

$$E_0 = \frac{\hbar^2}{2m}(k_{Ny})^2$$

$$E_p = \frac{\hbar^2}{2m}(B_{EC}k_{Ny})^2$$

$$V = E_0(k_{Ny}) - E_p(k_{Ny}) = \text{Potential.}$$

$$B_{EC} = \sqrt{\frac{k_{Ny}^2 - k_p^2}{k_{Ny}^2}}$$

Fig. 5. Relation between potential energy and momentum in classical model.
A classical analogy illustrating the relationship between momentum energy and the effective potential. In this model, the presence of a potential reduces the kinetic energy compared with the free-particle case, especially at the Nyquist wavenumber, where the kinetic energy is the maximum. Crucially, the potential here is not defined as a spatial function V(x), but as an effective quantity derived from the observed energy reduction at specific wavenumbers. This conceptual model anticipates the key argument of Section III: since the Bloch wave is physically meaningful only at discrete lattice positions, the detailed shape of V(x) between these points has no observable consequence. The energy diagram thus reflects only the net effect of the potential as encoded in the sampled modulation wave.

We determined the curvature coefficient $B_{EC}$ of the parabolic energy dispersion such that the resulting quadratic curve satisfies the two reference points, as follows:

$$B_{EC} = \sqrt{\frac{E_p(k_{Ny})}{E_0(k_{Ny})}} = \sqrt{\frac{\left(k_{Ny}^2 - \frac{2m}{\hbar^2}V\right)}{k_{Ny}^2}} = \sqrt{\frac{(k_{Ny}^2 - k_p^2)}{k_{Ny}^2}},$$

$$k_p^2 = \frac{2m}{\hbar^2}V \quad (13)$$

The energy of the Bloch wave in the presence of the potential can thus be expressed using the coefficient $B_{EC}$ as follows:

$$E_p(k) = \frac{\hbar^2}{2m}(B_{EC} \times k)^2 \quad \text{or} \quad E_p(k) = \frac{\hbar^2}{2m}(k_n \pm B_{EC} \times k)^2: k_n = \frac{2n\pi}{a} \quad (14)$$

($n$ is an integer that includes 0; $k_n$ is not affected by the potential)

This quadratic form reflects the modified curvature of the dispersion relation owing to the influence of the periodic potential.

### C. Comparison between the present analytical model and perturbation theory

The equations derived thus far fully describe the state of a 1D crystal lattice under a periodic potential. Therefore, no additional theoretical input is required to calculate the band gap. The band gap is defined as the energy difference between the maximum energy of the $n = 0$ band and minimum energy of the $n = 1$ band at $k = k_{Ny}$.

Thus, from Eq. (14), the maximum energy is given by

$$E_{n=0}(k_{Ny}) = \frac{\hbar^2}{2m}(B_{EC} \times k_{Ny})^2 = \frac{\hbar^2}{2m}\frac{(k_{Ny}^2 - k_p^2)}{k_{Ny}^2}k_{Ny}^2 = E_0 - V: n = 0 \quad (15)$$

The corresponding minimum energy in the $n = 1$ band is

$$E_{n=1}(k_{Ny}) = \frac{\hbar^2}{2m}(k_1 - (B_{EC} \times k_{Ny}))^2 = \frac{\hbar^2}{2m}(2k_{Ny} - (B_{EC} \times k_{Ny}))^2 \quad (16)$$

where $k_1 = \frac{2\pi}{a} = 2k_{Ny}$.

Fig. 6 presents the energy diagram based on Eqs. (15) and (16), comparing the cases with and without the potential. Accordingly, the band gap value can be expressed by Eqs. (17) and (18) as the energy difference relative to the case without the potential. As expected, the band gap $\Delta E_-$ is given by the difference between the energies at the Brillouin zone edges, as expressed in Eqs. (17) and (18).

$$\Delta E_- = E_{n=0}(k_{Ny}) - E_0 = (E_0 - V) - E_0 = -V \quad (17)$$

$$\Delta E_+ = E_{n=1}(k_{Ny}) - E_0 = \frac{\hbar^2}{2m}(2k_{Ny} - (B_{EC} \times k_{Ny}))^2 - \frac{\hbar^2}{2m}(k_{Ny})^2 \approx V \quad (18)$$

Eq. (17) is straightforward and easy to understand; in contrast, Eq. (18) appears more complex and less intuitive because it represents the exact solution, which—unlike in the perturbative approach—is not simplified. Therefore, we introduce an approximation for Eq. (18) to compare it with the result obtained from perturbation theory for $B_{EC}$. Assuming that $k_p \ll k_{Ny}$, we expand $B_{EC} = \sqrt{1 - \frac{k_p^2}{k_{Ny}^2}}$ by the Maclaurin series, and we obtain the following approximation: $B_{EC} \cong 1 - \frac{1}{2}\frac{k_p^2}{k_{Ny}^2}$

Then, $E_{n=1}(k_{Ny}) \approx \frac{\hbar^2}{2m}(2k_{Ny} - (B_{EC} \times k_{Ny}))^2 = \frac{\hbar^2}{2m}\left\{2k_{Ny} - \left(1 - \frac{1}{2}\frac{k_p^2}{k_{Ny}^2}\right) \times k_{Ny}\right\}^2.$

$$E_{n=1}(k_{Ny}) \approx \frac{\hbar^2}{2m}\left(k_{Ny} + \frac{1}{2}\frac{k_p^2}{k_{Ny}}\right)^2 = \frac{\hbar^2}{2m}\left(k_{Ny}^2 + k_p^2 + \frac{1}{4}\frac{k_p^4}{k_{Ny}^2}\right) \cong E_0(k_{Ny}) + V \quad (19)$$

This result coincides with the prediction from perturbation theory

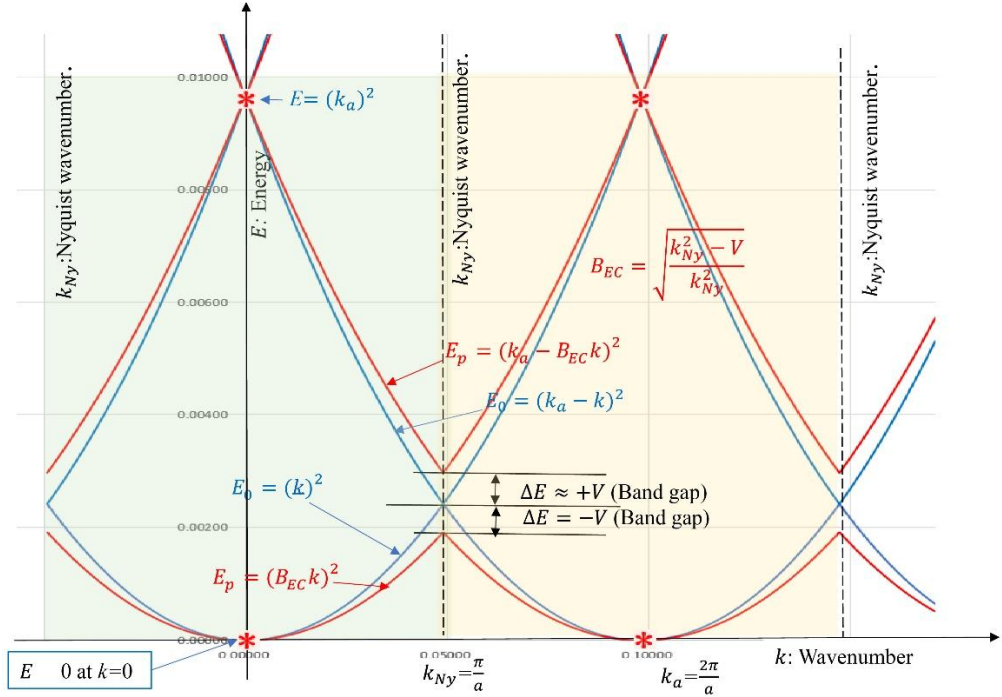

Fig. 6. Energy–wavenumber relation of the Bloch wave, showing the band gap formation due to the periodic potential. $k_a = \frac{2\pi}{a}$, $k_{Ny} = \frac{\pi}{a}$  $V$ = Potential, Energy, Wavenumber (Arbitrary unit).

## VI. Conclusion

This study explored the physical and intuitive mechanism behind band gap formation and provided direct mathematical support for this mechanism. Although further investigation may be needed to rigorously validate this approach, the recognition gained through this work, particularly regarding the role of discrete space and the act of observation, may offer broader insights into other physical phenomena.

**Acknowledgments**
The authors confirm that the data supporting the findings of this study are available within the article. The author would like to express sincere gratitude to Editage for English language editing and to OpenAI's ChatGPT for assistance in improving the linguistic clarity of the manuscript. Neither service contributed to the scientific content, data interpretation, or the creation of figures and tables.